# Quantum Capacitance-Limited MoS$_2$ Biosensors Enable Remote Label-Free Enzyme Measurements

Son T. Le[1,2], Nicholas B. Guros[3,4], Robert C. Bruce[1], Antonio Cardone[5,6], Niranjana D. Amin[7], Siyuan Zhang[1,2], Jeffery B. Klauda[4], Harish C. Pant[7], Curt A. Richter[1] and Arvind Balijepalli[3,*]

[1]Nanoscale Device Characterization Division, National Institute of Standards and Technology, Gaithersburg, MD 20899, USA ; [2]Theiss Research, La Jolla, CA 92037; [3]Microsystems and Nanotechnology Division, National Institute of Standards and Technology, Gaithersburg, MD 20899, USA; [4]Department of Chemical and Biomolecular Engineering, University of Maryland, College Park, MD 20742, USA; [5]Software and Systems Division, National Institute of Standards and Technology, Gaithersburg, MD 20899, USA; [6]University of Maryland Institute for Advanced Computer Studies, University of Maryland, College Park, MD 20742, USA; [7]National Institute of Neurological Disorders and Stroke, National Institutes of Health, Bethesda, MD 20892, USA
*e-mail: arvind.balijepalli@nist.gov

**Abstract:**
We have demonstrated atomically thin, quantum capacitance-limited, field-effect transistors (FETs) that enable the detection of pH changes with ≈75-fold higher sensitivity (4.4 V/pH) over the Nernst value of 59 mV/pH at room temperature when used as a biosensor. The transistors, which are fabricated from a monolayer of MoS$_2$ with a room temperature ionic liquid (RTIL) in place of a conventional oxide gate dielectric, exhibit very low intrinsic noise resulting in a pH limit of detection (LOD) of $92 \times 10^{-6}$ at 10 Hz. This high device performance, which is a function of the structure of our device, is achieved by remotely connecting the gate to a pH sensing element allowing the FETs to be reused. Because pH measurements are fundamentally important in biotechnology, the low limit of detection demonstrated here will benefit numerous applications ranging from pharmaceutical manufacturing to clinical diagnostics. As an example, we experimentally quantified the function of the kinase Cdk5, an enzyme implicated in Alzheimer's disease, at concentrations that are 5-fold lower than physiological values, and with sufficient time-resolution to allow the estimation of both steady-state and kinetic parameters in a single experiment. The high sensitivity, low LOD and fast turnaround time of the measurements will allow the development of early diagnostic tools and novel therapeutics to detect and treat neurological conditions years before currently possible.

Rapid and sensitive pH measurements based on field-effect transistors (FETs) are used in diverse applications that include determining the effects of ocean acidification on marine ecology,[1] biomanufacturing,[2] and low cost DNA sequencers.[3] However, drastic improvements in sensitivity and the limit of detection (LOD) of electronic pH transduction are needed to accelerate their widespread use in important biotechnology applications. One such example is the measurement of the function of enzymes, macromolecular catalysts that accelerate biochemical reactions and serve an integral role in ensuring normal cellular function,[4] where small changes in the solution pH act as a reporter of their function.[5,6] Disruptions in normal enzyme function are known to give rise to debilitating neurological conditions including Alzheimer's and Parkinsons's disease,[7] several cancers,[8] and even chronic neuropathic pain.[9] As part of normal cellular function, enzymes catalyze the phosphorylation of substrate proteins by the hydrolysis of adenosine triphosphate (ATP) and the transfer of a single phosphate group. This results in the release of one or more protons into solution, and thereby a small change in the pH. The change in pH typically varies by less than 0.005 units[10,11] under physiological conditions (i.e., for normal living organisms) or an order than magnitude lower than the resolution demonstrated with state-of-the-art ion-sensitive field effect transistors (ISFETs). [3,12]

Several techniques have been developed over the past century for more accurate measurements of solution pH. The Harned cell is the primary pH measurement standard,[13] but it requires an elaborate setup and long equilibration times making it unusable in the measurement of small and fast pH changes. As a result, pH measurements used in biotechnology applications rely extensively on electrode measurements or spectrophotometric techniques. These approaches lack adequate sensitivity and resolution for accurate measurements of enzyme mediated pH changes.[14,15] In the past several years, there have been numerous developments in pH sensing technology using solid state devices that are potentially suitable for biological applications.[3,12,16,17] Improved sensitivity and dynamic measurements have been demonstrated with ISFETs[3] and more recently dual-gate silicon FETs.[16] In the latter case, pH sensitivity was shown to be approximately two-fold higher than the Nernst potential of 59 mV/pH at room temperature.[17] Notably, the LOD of ISFETs has been reliably demonstrated to be 0.05 pH units.[3,12] The LOD for these devices could be lower if the device parameters are optimized[18] or if the measurements are performed under a very narrow bandwidth.[19] However, resolution that is adequate for enzyme measurements under physiological conditions are yet to be experimentally realized with silicon FETs. These limitations have required FET-based enzyme catalyzed phosphorylation assays to be performed at concentrations that are about two orders of magnitude higher than physiological values, precluding their use in diagnostic and therapeutic development applications.[6,20,21]

Here, we show that FETs with an atomically thin semiconducting channel and gated with an ionic liquid (2D-ILFET) allow label-free measurements of enzyme activity and kinetics at or below physiological concentrations. The measurements were performed by using a pH sensing element that is connected electrically to the 2D-ILFET (Figure 1a). This configuration allowed measurements with both a high intrinsic gain[16] (i.e., sensitivity) and signal-to-noise ratio (SNR) that exceeded comparable silicon devices by more than an two orders of magnitude.[22] We show that the newly-developed 2D-ILFET measurements are a quantitative first step to enabling their use in drug discovery and diagnostic applications. The ability to perform measurements within current testing frameworks without resorting to surface modification[5,20,23] or hazardous radioactive labeling will enable rapid adoption of the method. Furthermore, the higher sensitivity of the technique will allow its use in applications beyond enzymatic measurements, for example in critically needed early detection of neurodegenerative conditions[24] and more broadly to the measurements of biomolecular interactions that are important in many applications in biophysics and biotechnology.[3,25-27]

In order to optimize the use of ionic liquid-gated 2D-ILFETs in enzyme activity measurements, we first characterized their electrical performance, followed by sensitive pH measurements to validate the approach. The optimized configuration was then used to measure the activity and kinetics of the proline directed kinase Cdk5,[28] an enzyme strongly implicated in several neurodegenerative conditions.

**Methods**

**Device Fabrication and Characterization:** Large area, high quality monolayer $MoS_2$ flakes were exfoliated from bulk crystal using the gold-mediated exfoliation method[29] on to oxidized, heavily doped silicon substrates that also served as a global back-gate. $SiO_2$ substrates with a thickness of 300 nm and 70 nm were used to study the device performance scaling. After the monolayer $MoS_2$ flakes were located and inspected using an optical microscope, optical photolithography was used to pattern the source, drain, and gate electrodes followed by electron beam evaporation of metal Ti/Au (2 nm/100 nm) and lift-off in acetone. A second photolithography step was used to define the active channel area, followed by reactive ion etching to remove excess $MoS_2$. After a final resist cleaning, a small droplet of the DEME-TFSI IL (727679; Sigma Aldrich, St. Louis, MO) was carefully applied onto the devices using a micromanipulator under an optical microscope. The droplet was sized to cover the $MoS_2$ monolayer and the gate electrodes,[30,31] as seen in the optical image of the final device in Figure 1b. It is important to note that for good ionic liquid-gate coupling with the device's channel, the area of the gate electrodes that is in contact with the ionic liquid was designed to be much larger than the combined area of the source, drain contacts, and $MoS_2$ channel that are in contact with the ionic liquid. Gate leakage was verified by measuring the current across the ionic liquid upon application of a potential across two patterned gate electrodes. The measurement scheme for electrical characterization is shown Figure 1a.

**Quantum Capacitance Model:** For single layer $MoS_2$ FET under positive ionic liquid-gate bias, the relationship between gate voltage, $V_{LG}$, and the channel charge carrier density, $n_{ch}$, is given by[32,33]

$$V_{LG} - V_{t,LG} - V_{FB} = \frac{E_g}{2q} - \frac{k_B T}{q} \ln\left[\exp\left(\frac{n_{ch}}{g_{2D} k_B T}\right) - 1\right] + \frac{q n_{ch}}{C_{LG}} \quad (1)$$

where $V_{t,LG}$ is device threshold voltage, $V_{FB}$ is the flat band voltage, $Eg$ is material band gap, $q$ is elementary charge, $k_B$ is Boltzalmann's constant, $T$ is the temperature in Kelvin, $g_{2D}$ is the band edge density of states, and $C_{LG}$ is the liquid-gate capacitance. For a fixed $V_{LG}$, we calculated $V_{t,LG}$ as a function of channel carrier density.

$$V_{t,LG} = V_{LG} - V_{FB} - \left\{\frac{E_g}{2q} - \frac{k_B T}{q} \ln\left[\exp\left(\frac{n_{ch}}{g_{2D} k_B T}\right) - 1\right] + \frac{q n_{ch}}{C_{LG}}\right\} \quad (2)$$

The channel quantum capacitance ($C_Q$) is dependent on the density of charge carriers within the channel and is given by,[32]

$$C_Q \approx q^2 g_{2D} \left[1 + \frac{\exp\left(\frac{E_g}{2k_B T}\right)}{2\cosh\left(\frac{q V_{ch}}{k_B T}\right)}\right]^{-1}, \quad (3)$$

where

$$V_{ch} = \frac{E_g}{2q} - \frac{k_B T}{q} \ln\left[\exp\left(\frac{n_{ch}}{g_{2D} k_B T}\right) - 1\right] \quad (4)$$

Using Equations 1—4 we can calculate $V_{t,LG}$ and $C_Q$ as a function of $n_{ch}$. Since $C_{TG}$ and $C_Q$ are in series as seen from Figure 1e, we calculate $C_{TG}(C_Q) = C_{LG} \times C_Q/(C_{LG}+C_Q)$, where $C_{LG}$ is the ionic liquid-gate capacitance. Similarly, the back-gate capacitance, $C_{BG} = C_{ox} \times C_Q/(C_{ox}+C_Q)$, where $C_{ox}$ is the back-gate oxide capacitance. Moreover, since $C_Q \gg C_{ox}$, $C_{BG} \approx C_{ox}$=constant. Under these limits, the expression for $\alpha$ is given by,

$$\alpha = \frac{C_{TG}}{C_{BG}} \approx \frac{C_{LG} C_Q}{C_{ox}(C_{LG}+C_Q)}, \qquad (5)$$

and is limited by $C_Q$. At large values of $n_{ch}$, $C_Q$ approaches its theoretical limit, $C_{Q,max}$, and $\alpha$ is constant. As $n_{ch}$ decreases, $C_Q$ decreases exponentially from $C_{Q,max}$ and we observe a corresponding reduction in $\alpha$. The results of the model were compared with experimentally measured values of $\alpha$, which were extracted by taking the numerical derivative of $V_{BG}$ ($V_{t,LG}$) (Figure 1d and Figure S2; *Supplementary Information*).

**Time-Series Field-Effect Transistor Measurements and PID Control:** Time-series measurements were performed following the schematic in Figure 2a. The signal on the ionic liquid-gate ($V_{LG}$) was switched between an arbitrary function generator (HF2LI; Zurich Instruments, Zurich, Switzerland) or a pH microelectrode (MI-4156; Microelectrodes, Bedford, NH). An offset voltage $V_o$ (GS200; Yokogawa Corporation, Tokyo, Japan) was then added to $V_{LG}$ using a summing amplifier (SR560; Stanford Research Systems Inc., Sunnyvale, CA). The FET was operated in a constant current mode using a PID controller that varied $V_{BG}$ in response to changes in $I_D$. The channel current was first amplified using a current preamplifier (DLPCA-200; FEMTO, Berlin, Germany) with a transimpedance gain of $10^6$ V/A. The amplified voltage was input to a digital PID controller (HF2LI; Zurich Instruments, Zurich, Switzerland), filtered using a 4-pole Bessel low pass filter (LPF) with a bandwidth of 5 kHz and then sampled at 25 kHz using a 14-bit analog to digital converter. The PID controller varied $V_{BG}$ in response to changes in $I_D$ with a bandwidth of 1 kHz ($K_P$=496.1, $K_I$=9.242×10$^3$ s$^{-1}$ and $K_D$=8.02 μs). The PID output was allowed to vary between -10 V to +10V to drive the back-gate voltage ($V_{BG}$).

**Ionic Liquid-Gate C-V Measurements:** In order to verify the capacitance ($C_{LG}$) of the ionic liquid, we utilized a LCR meter (E4980A; Agilent, Santa Clara, CA) to make 2-probe, AC capacitance measurements across the ionic liquid, measuring across a frequency range of 20 Hz to 2 MHz. Figure S3 (see *Methods*) shows representative results of these capacitive measurements, highlighting that at low frequency (where our device operates for sensing) the capacitance constant and approximately 1 nF (see Figure S3, *inset; Methods*). As seen in the inset, we observe a negligible bias dependence of the capacitance at all measured frequencies. The capacitance exhibits a less than 1% change over a ± 0.5 V range at the lowest frequency (20 Hz). Due to the design of our devices, we do not have a constrained contact area of the ionic liquid on electrodes, therefore we utilized optical estimates of ≈100 μm × 100 μm as the area of metal plates that are in contact with the ionic liquid used in the capacitance measurement. This allowed us to estimate the specific gate capacitance to the liquid-gate to be ≈10.7 μF/cm$^2$ at low frequency range (20 Hz to 2 kHz), in good agreement with values published in the literature.[34]

**Kinase Measurement Reagents:** Static activity measurements of Cdk5/p25 phosphorylation of histone H1 were performed by suspending 100 ng of the enzyme (C0745; Sigma Aldrich, MO) in 50 μL of 1× kinase buffer to obtain a final concentration of 18.5 nM. The Cdk5/p25 concentration was reduced five-fold for dynamic measurements to obtain a final concentration of 3.7 nM. A stock solution of 5× kinase buffer was prepared by suspending 25 mM β-glycerol (G9422; Sigma Aldrich, MO), 50 mM MgCl$_2$ (5980; Millipore, MA), 5 mM EGTA (E0396; Sigma Aldrich, MO), 2.4 mM EDTA (1002264786; Sigma Aldrich, MO), 1.25 mM MOPS (M1254; Sigma Aldrich, MO) in deionized water (DIW) and diluting further to form 1× kinase buffer. Solutions of the substrate protein were prepared by adding 2 mg/mL of histone H1 (10223549001; Sigma Aldrich, MO) to the assay to obtain the final concentrations as described in the main text. The phosphorylation reaction was triggered with a cocktail of DTT and ATP. The final concentration of the ATP and DTT solution was adjusted to 250 μM and 5 mM in DIW respectively.

**Cdk5 γ-$^{32}$P-ATP Measurements:** Cdk5/p25 (C0745; Sigma Aldrich, MO) kinase activity was measured as described in the manufacurer's protocol with some modifications. The kinase reaction was initiated by adding γ-$^{32}$P-ATP (final concentration of 50 μM) to a preincubated substrate buffer cocktail and incubated at 30 $^0$C for 1 hour. The reaction was terminated by spotting 20 μl of the reaction mixture on a P81 phosphocellulose pad. Dried pads were washed 3 times in 75 mM phosphoric acid, followed by rinsing with 95% ethanol. The radioactivity of the spoted pads was measured in a liquid scintillation counter. Appropriate controls, without added phosphoryl acceptor substrates, were also run and the values were subtracted from the total counts obtained in the presence of substrate protein.

**Results**

**Ionic Liquid-Gate 2D FET Performance.** Initial characterization of 2D-ILFETs fabricated with monolayer MoS$_2$[35] was performed with the set up in Figure 1a and using a representative device seen in Figure 1b (see *Methods* for fabrication details) where the sensing signal was connected to the top ionic liquid gate, while the silicon substrate was used as the measurement gate. The devices described in this work are fundamentally different from dual-gate silicon FETs[16,36] in that the asymmetric, capacitively coupled, gates in the 2D-ILFET control the same atomically thin semiconducting channel giving rise to a large intrinsic gain in conjunction with ultralow noise performance. Furthermore, the measurement setup allowed the 2D-ILFET performance to approach the intrinsic quantum capacitance limit of the channel material.

The transfer characteristics of the device were measured by recording the drain current ($I_D$) as a function of the liquid-gate potential ($V_{LG}$) when the drain voltage ($V_D$) was held constant. The measurements were repeated for different back-gate voltages ($V_{BG}$) to estimate the signal amplification ($\alpha$) due to the asymmetric capacitance of the ionic and back gates (Figure 1c). The devices exhibited a dynamic range of ≈5 orders of magnitude in $I_D$ and a subthreshold slope between 69 mV/dec and 145 mV/dec at ≈300 K.

For each curve in Figure 1c, the liquid-gate threshold voltage ($V_{t,LG}$) was estimated from a linear extrapolation of the peak transconductance to the x-axis.[37] The value of $\alpha = dV_{BG}/dV_{t,LG}$ for representative devices was then determined numerically from the data in Figure 1d ($d$=300 nm) and Figure S2 ($d$=70 nm; see *Supplementary Information*), where $d$ is the thickness of the back-gate dielectric. From Figure 1d we are able to clearly discern two distinct regimes for $V_{BG}$ as a function of $V_{t,LG}$. At large and positive $V_{BG}$ (Figure 1d; *pink*), where the device was in the limit of large 2D channel carrier density ($n_{ch}$), $\alpha$ was constant and estimated to be 156±3 ($d$=300 nm) and 41±4 ($d$=70 nm). The error bars report the standard error. On the other hand, at negative $V_{BG}$ and low $n_{ch}$ (Figure 1d; *blue*), $\alpha$ decreased exponentially. Both regimes of operation are explained by the quantum capacitance model discussed next.

**Quantum Capacitance-Limited Performance.** For devices with two gates, $\alpha$ can also be determined from the ratio of the top ($C_{TG}$) and bottom gate ($C_{BG}$) interface capacitances, i.e., $\alpha \approx C_{TG}/C_{BG}$.[16,30,38] For dual-gate FETs where the individual gate oxide capacitances are considerably lower than those associated with the channel, $\alpha$ is constant and can be estimated directly from the properties of the gate dielectric.[34] However, due to the large capacitances associated with ionic liquids, for the devices reported here $C_{TG}$ is limited by the intrinsic quantum capacitance ($C_Q$) of the 2D semiconducting channel. Importantly, $C_Q$ approaches its theoretical limit, $C_{Q,max}$ when $n_{ch}$ is large.[32] Under these conditions $\alpha$ is constant. On the other hand, when $n_{ch}$ is reduced, $C_Q$ and therefore $\alpha$ decrease exponentially (see *Methods* for a detailed explanation). This transition between the linear and exponential regimes can be observed in terms of the gate voltage in Figure 1d and $\alpha$ in Figure 1f.

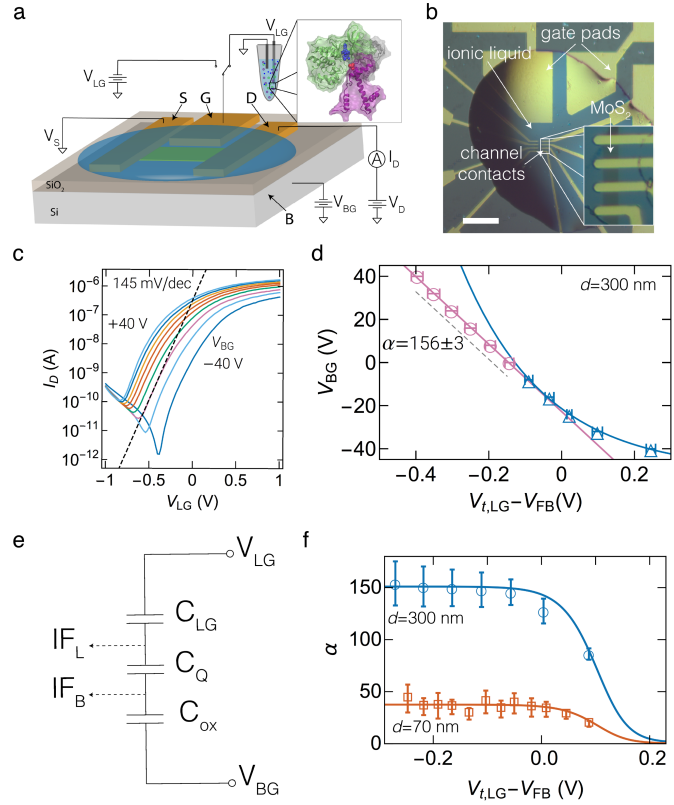

Figure 1: Electrical characterization of ionic liquid-gate field-effect transistors (FET). Error bars report standard errors. (a) Device schematic of an ionic liquid-gated FET for biosensing. A channel formed between the source (*S*) and drain (*D*) terminals is controlled electrostatically by a voltage applied to the silicon substrate (*B*) or the ionic liquid top-gate (*G*). A voltage applied to the ionic liquid-gate ($V_{LG}$) can be switched between a voltage source for characterization or a biosensing element. (b) An array of ionic liquid dual-gate FETs fabricated using a 2D $MoS_2$ film on a 300 nm $SiO_2$ substrate. The scale bar represents 100 μm. (c) Transfer characteristics show the drain current ($I_D$) as a function of $V_{LG}$ and varying back-gate voltage ($V_{BG}$). (d) The change in $V_{BG}$ as a function of liquid-gate threshold voltage ($V_{t,LG}$). A flat band voltage ($V_{FB}$) of -0.82 V was subtracted from $V_{t,LG}$ to enable comparison with theory. (e) The dual-gated FET was modeled with three capacitors in series. The quantum capacitance ($C_Q$) of the 2D channel controls the capacitive coupling between the back ($IF_B$) and liquid ($IF_L$) gate interfaces. (f) A plot of $\alpha$ as a function of $V_{t,LG}$ for two devices with back-gate oxide thickness of 300 nm (○) and 70 nm (□). The solid line shows the theoretical model for quantum capacitance limited device performance. $V_{FB}=-0.82$ V ($d=300$ nm) and $V_{FB}=+0.12$ V ($d=70$ nm) were subtracted from $V_{t,LG}$ to allow direct comparison of the devices.

We see quantitative agreement between the model and measurements for two representative device structures shown in Figure 1f. In both cases, we assumed $C_{ox}$=0.0115 µF/cm² ($d$=300 nm) and 0.049 µF/cm² ($d$=70 nm), and $C_{LG}$=10.7 µF/cm² based on our own measurements (Figure S3; see *Supplementary Information*) and literature values.[34] $C_{Q,max}$ was directly calculated from Equation 5 (see *Methods*) to be 2.16 µF/cm², and was in excellent agreement with the value of 2.2±0.05 µF/cm² extracted from a non-linear regression of the model to the data in Figure 1f. The only other free parameter in the model was the flat band voltage ($V_{FB}$), which depends in part on the fabrication process.[39] The extracted value of $C_{Q,max}$ was within ≈45 % of the maximum theoretical value for monolayer MoS$_2$, and more than two times higher than previous measurements.[34,40] This in turn allowed the devices to operate with high sensitivity when used for enzyme measurements. Furthermore, the ability to tune $\alpha$ with only the gate voltage is advantageous in biosensing applications to allow sensitivity to be offset for higher dynamic range.

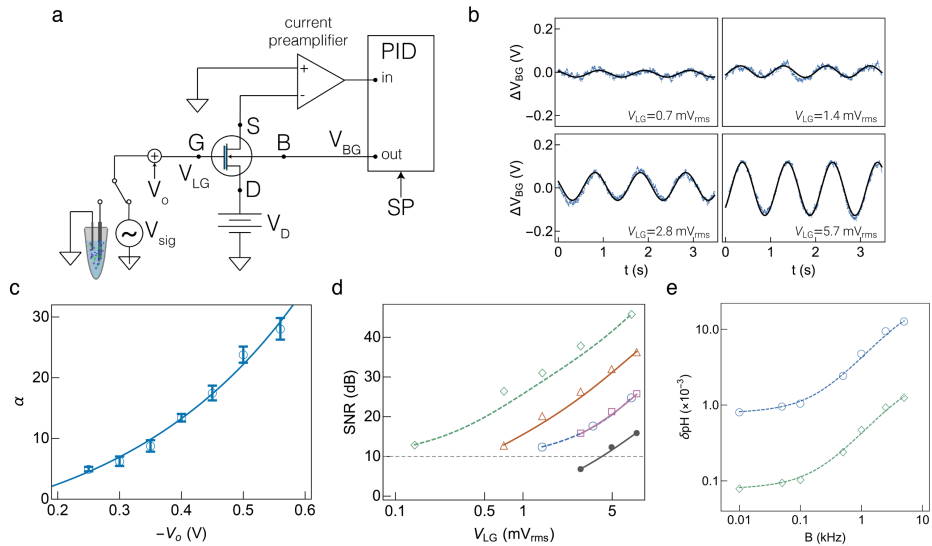

Figure 2: Constant current mode electrical characterization of dual-gated ionic liquid-gated field-effect transistors (FET). (a) The FETs were set up in a constant current mode using a proportional-integral-derivative (PID) controller. $I_D$ was held constant by continually adjusting $V_{BG}$ in response to small changes to $V_{LG}$. (b) The response of $V_{BG}$, under PID control, as a function of time is shown upon application of a 1 Hz AC sine wave signal with varying amplitude to the ionic liquid-gate. Data are shown for a back-gate

oxide thickness, $d$=300 nm. (c) The amplification at the back-gate ($\alpha$) increased with the applied liquid-gate offset voltage, $V_o$, allowing the device gain to be smoothly tunable. The error bars report the standard error. (d) The signal-to-noise ratio (SNR) was estimated with a bandwidth of 5 kHz as a function of AC signal amplitude for devices operated under PID control and in open loop (-●-). For devices fabricated with either 300 nm ($\alpha$ =13; ⊟ and $\alpha$=42; ▲) or 70 nm ($\alpha$ =5; ⊖ and $\alpha$ =50; ⬦) SiO$_2$ back-gate, the SNR under PID control was higher than under open loop operation. (e) The limit of pH detection as a function of the measurement bandwidth and $\alpha$ for devices with a 70 nm ($\alpha$ =5; ⊖ and $\alpha$ =50; ⬦) SiO$_2$ back-gate dielectric.

**Constant Current Operation for Biosensing.** The preceding results were used to maximize $\alpha$ when operating the ionic liquid FETs in a constant current mode using a proportional-integral-derivative (PID) controller as shown in Figure 2a. The PID controller varied $V_{BG}$ in response to $V_{LG}$ to maintain a root mean square (RMS) channel current of 100 nA. PID performance was then compared with open loop operation where $I_D$ was recorded in response to changes in $V_{LG}$. The signal connected to the ionic liquid-gate was setup to allow switching between an arbitrary waveform generator to calibrate sensor performance or a biosensing element. In both cases, a fixed DC offset voltage ($V_o$) was added to $V_{LG}$ to set the value of $\alpha$ (see *Methods*).

**Device Sensitivity.** Device calibration was performed with a 1 Hz sine wave applied to the ionic liquid-gate. Figure 2b shows the change in $V_{BG}$ ($d$= 300 nm) under PID control with $V_o$=–0.56 V for sine wave amplitudes of $V_{LG}$ ranging from 0.7 to 5.7 mV$_{rms}$, which resulted in $\alpha$=28. Tuning $V_o$ allowed sensitivity to be smoothly offset for dynamic range with the highest $\alpha$ realized when operating near the linear regime determined from Figure 1d. Using this approach, we tuned $\alpha$ smoothly from 5±0.4 to 28±1.8 (Figure 2c) when $d$=300 nm or from 5±0.5 to 50±1.5 when $d$=70 nm. Moreover, the measured values of $\alpha$ were consistent with those in Figure 1f, and more than an order of magnitude higher than dual-gate silicon FET measurements.[16]

**Signal-to-Noise Ratio and Limit of Detection.** The signal-to-noise ratio (SNR) is a true measure of sensor performance. To determine if the higher sensitivity of our FETs, relative to dual-gate silicon devices,[16] translated to an improved LOD, we measured the noise in $V_{BG}$ and $I_D$ of the 2D-ILFET devices as shown in Figure S4. These measurements were then used to estimate the SNR of the FET in PID and open loop modes. Figure S4 shows a representative power spectral density (PSD) of the back-gate voltage (*top*) under PID control ($d$=300 nm) and channel current (*bottom*) during open loop operation. The broadband noise under PID control was estimated using the expression $\delta V_{BG} = \sqrt{\int_{BW} S_{V_{BG}} df}$, from DC to the low pass filter bandwidth of 5 kHz and found to be 5.8 mV$_{rms}$ ($d$=300 nm; Figure S4 *top*) and 1.8 mV$_{rms}$ ($d$=70 nm; Figure S4 *middle*) for the measured devices, decreasing as expected with the back-gate oxide thickness. Furthermore, $\delta V_{BG}$ was found to be invariant with $V_o$ and thereby $V_{BG}$. The channel current noise in the open loop case, $\delta I_D = \sqrt{\int_{BW} S_{I_D} df}$ was 700 pA$_{rms}$ (Figure S4 *bottom*). The SNR was estimated using the expressions 20 $log_{10}(V_{BG}/\delta V_{BG})$ under PID control and 20 $log_{10}(I_d/\delta I_d)$ in open loop mode and shown in Figure 2d as a function of $V_{LG}$. For a bandwidth of 5 kHz, SNR was found to increase ≈3-fold ($\alpha$=13, $d$=300 nm) to ≈30-fold ($\alpha$=50, $d$=70 nm) under PID control when compared to open loop operation.

The LOD of the devices under PID control was estimated as a function of the measurement bandwidth when SNR=3 (or ≈10 dB). This in turn allowed the estimation of the pH resolution ($\delta$pH) at a temperature of 300 K as seen from Figure 2e. At a bandwidth of 5 kHz, LOD was estimated to be 768 μV ($\alpha$=5, $d$=70 nm) resulting in $\delta$pH=0.013 and only 77 μV ($\alpha$=50, $d$=70 nm) or $\delta$pH=0.0013. Since the measurements of the kinetics of biomolecules are relatively slow, $\delta$pH can be further decreased at lower bandwidth. When the signal was measured with a bandwidth of 10 Hz (Figure 2e; $\alpha$=50, $d$=70 nm), we estimated $\delta$pH=82×10$^{-6}$, representing a greater than 10-fold improvement over state-of-the art ISFET measurements.[1,12,18,19] Importantly, in contrast to measurements made with dual-gate Si FETs, the LOD improved with increasing $\alpha$ as discussed further below.

The device structures described here are fundamentally different from silicon-based dual-gated FETs described in previous studies.[16,36] In those devices, it was interpreted that two channels were formed and controlled independently by the top and bottom gates. In constant current mode, when the device was operated in the inversion-inversion regime, changes in the gate potential at one interface that drive the corresponding channel into strong inversion (increasing channel current), cause the other channel to be placed into weak inversion (with higher noise). Regardless of the polarity of the applied potential at the sensing gate, the overall channel current noise is dominated by the channel in weak inversion. Channel noise could be further limit SNR when the devices are operated in the inversion-depletion regime, for example as demonstrated with ultra-thin body double-gated silicon FETs.[16] In this case, the noise is dominated by the depleted channel and should result in lower performance compared to the inversion-inversion regime. Furthermore, this effect becomes more pronounced with increasing gain. As a result, extensive work with such devices has shown no measurable improvement in SNR of silicon devices despite large improvements in the sensitivity.[36] In contrast, ultra-thin channels formed with 2D materials have a single conducting channel controlled by both the top ionic liquid-gate and bottom $SiO_2$ back-gate. Because the channel can always be placed in the inversion regime, the relative noise is low over a wide range of biasing conditions leading to improved SNR with these devices.

**Sensitive pH Measurements.** The ionic liquid-gate FETs were used to remotely measure the pH of buffered solution with high sensitivity. In this measurement, $I_D$ was measured as a function of $V_{BG}$ when a pH sensor measuring standard buffer solutions was connected to the liquid-gate as seen in Figure 3a (*inset*). The change in the back-gate threshold voltage, $V_{t,BG}$ for each measured pH from 4 to 10 is apparent from the data in Figure 3b and was found to be linear (Figure 3a) as a function of pH. The measurement sensitivity was estimated to be 4.4 V/pH or ≈75× higher than the Nernst limit of 59 mV/pH at room temperature[17] (Figure 3a; $\alpha$=159) and is a function of the device's intrinsic $\alpha$.

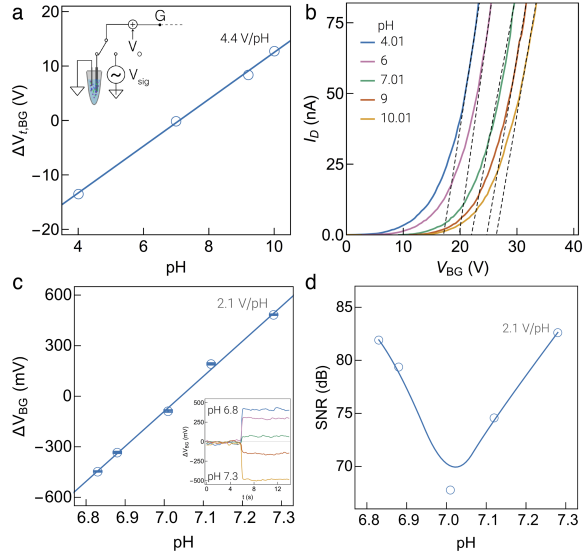

Figure 3: Ionic liquid-gate field-effect transistor (FET) calibration with buffered pH solutions. (a) The change in the back-gate threshold voltage ($\Delta V_{t,BG}$) as a function of the solution pH showed a linear response over the measured range. When using a 300 nm SiO₂ back-gate, the pH sensitivity was found to be 4.4 V/pH (○). (*inset*) Measurements of pH were performed by connecting the ionic liquid-gate (G) to a pH sensing element. The dashed line shows the connection to the FET. (b) pH sensitivity of buffered electrolyte solutions was extracted from the shift in $\Delta V_{t,BG}$. Measurements were performed over a wide range of solution pH from 4.01 to 10.01. (c) The back-gate voltage ($V_{BG}$) response to varying pH buffer solutions in constant current mode with a bandwidth of 10 Hz is shown when $\alpha$ =33 (⊖), resulting in a sensitivity of 2.1 V/pH. The error bars represent the standard error. (*inset*) Time-series measurements of $V_{BG}$ response to buffered pH solutions. (d) The measurement signal to noise ratio (SNR) was measured at a bandwidth of 10 Hz and found to average 75 dB (≈5600). SNR was lowest in the vicinity of neutral pH (≈7) where $V_{LG}$ was very small.

The solution pH was also measured in a constant current mode using PID control by setting $\alpha$=33 ($V_o$=-1.5 V, $d$=300 nm). Figure 3c (*inset*) shows time-series measurements of phosphate buffered saline (PBS) solution at a bandwidth of 10 Hz. A switch was used to alternatively connect the ionic liquid-gate to the PBS solution and to ground. Figure 3c plots the change in $V_{BG}$ with solution pH. Measurements were made relative to pH 7 ($V_{LG}$=0 V); acidic pH resulted in $V_{LG}$>0 V and basic pH in $V_{LG}$<0 V. A linear least-squares fit to the data returned a slope of 2.1 V/pH when $\alpha$=33. Similar to measurements in Figure 2, sensitivity can be tuned by varying $V_o$ to alter $\alpha$. For example, when $\alpha$=23 we obtained 1.2 V/pH, which could allow measurements over a larger range (*data not shown*). The maximum value of $\alpha$ obtained in the constant current mode was constrained by technical limitations in the control electronics, which precluded measurements at the maximum value of $\alpha$=159 for this device as seen from the static pH measurements in Fig. 3a.

Measurement noise was estimated to be ≈60 $\mu V_{rms}$ by integrating the PSD of VBG from DC to 10 Hz. The noise spectra when measuring pH did not change substantially from those shown in Figure 2. The SNR exhibited a minimum near pH 7 when $V_{LG}$ was small (≈ 68 dB at a sensitivity of 2.1 V/pH; Figure 3d). Based on the pH measurements in Fig. 3 and the noise spectra, the LOD of the devices ($\alpha$=33) were determined to be 92 × 10$^{-6}$ pH units at the noise floor (SNR 3; ≈10 dB) for a bandwidth of 10 Hz. Because the device LOD improves with $\alpha$ as seen in Fig. 2, the pH measurements can be further improved in constant current mode by increasing $\alpha$ to its theoretical limit of 159 ($d$=300 nm) and will be explored in future work.

**Enzyme Activity and Kinetics.** Protein kinases are an very important subset of enzymes that facilitate post translational modifications, such as the phosphorylation of substrate proteins, to influence signaling pathways in the cell cycle and achieve homeostasis.[4,41] The deregulation of protein kinases is believed to underlie the onset of neurological conditions such as Alzheimer's disease or Parkinson's disease.[42,43] Therefore, the functional quantification of enzyme activity is central to the development of therapeutics that restore physiological function.[44] Existing methods to measure enzyme activity, for example during protein phosphorylation, which rely on radioactively labeled adenosine triphosphate (ATP) analogs or fluorophores, are expensive because they need specialized molecule labeling and handling, require hours to yield results, and often alter the kinase activity, thus limiting their effectiveness in rapid therapeutic screening.[45,46]

We applied our ultra-sensitive FET devices to measure the activity and kinetics of the proline directed kinase Cdk5. Under normal physiological conditions, Cdk5 is tightly regulated by either the p35 or p39 inhibitory proteins.[7,47] Oxidative stress causes a 10 kDa membrane anchored fragment of p35 to be cleaved, forming the pathological activator p25, resulting in deregulation and delocalization of the complex to the cytosol.[48] The resulting pathological complex, Cdk5/p25, has higher activity than its physiological counterpart, Cdk5/p35, and participates in the indiscriminate phosphorylation of numerous proteins, which are known to result in neurofibrillary tangles that are a hallmark symptom of Alzheimer's disease.[49]

The multi-protein pathological complex, Cdk5/p25, participates in the phosphorylation reaction shown in Figure 4a. Cdk5 mediated phosphorylation results in the release of a proton during ATP hydrolysis and the transfer of a single phosphate group to either a serine or threonine residue immediately preceding a proline.[5] Here, we demonstrate the ability of ionic liquid-gated FETs to detect small changes in the solution pH during phosphorylation of the substrate protein, histone H1 (see *Methods* for reagent conditions).

**Enzyme Activity.** Figure 4b shows the change in $V_{t,BG}$ as a function of histone H1 concentration under steady-state conditions. In each case, $V_{t,BG}$ was estimated from an $I_D$-$V_{BG}$ plot shown in Figure S6 (see *Supplementary Information*). To account for instrument drift, each data point in the figure was measured differentially with a control sample that was identical to the measured vials except for the absence of ATP, thereby inhibiting the phosphorylation reaction. The kinase activity was then estimated using, $s = \gamma \frac{[H1]}{k+[H1]}$, where $k$ is a constant, [H1] is the concentration of histone H1 and $\gamma$ is a scaling constant. For the FET measurements in Figure 4b, we estimated $k$=17.5±1.3 µM from a non-linear regression of the model to the measured data, consistent with previously published activity measurements for the pathological Cdk5/p25 complex.[10] We compared the measurements in Figure 4b against enzyme activity measurements obtained from a radioactively labeled γ-$^{32}$P-ATP assay (see *Methods*) as seen in Figure 4c.[50] The estimated value of *k*=12.1±2.3 μM from those measurements was found to be statistically consistent with the FET measurements with 95 % confidence.

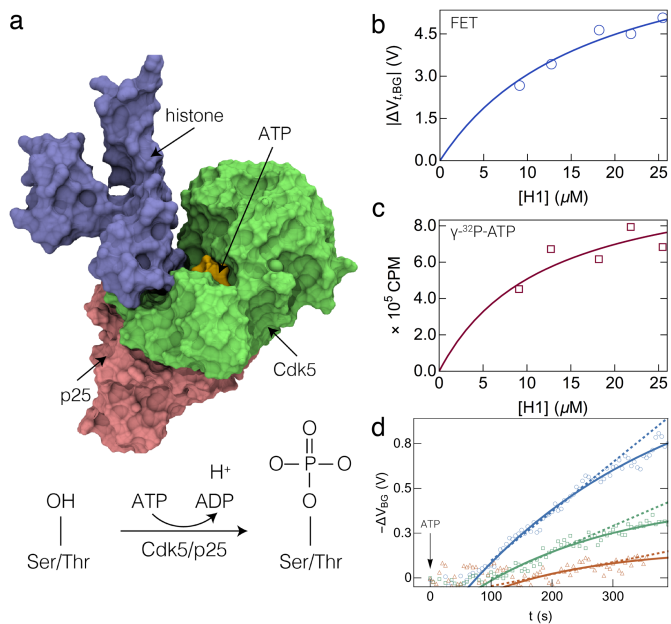

Figure 4: Ionic liquid-gate field-effect transistors (FET) were used for label-free enzyme activity and kinetics measurements. (a) The proline directed kinase Cdk5 catalyzes the phosphorylation of substrate proteins (e.g., histone H1) in the presence of an activator (e.g., p25) and adenosine triphosphate (ATP). The hydrolysis of ATP results in the transfer of a single phosphate group to either a serine (Ser) or threonine (Thr) residue in the substrate protein and the release of a proton into solution, resulting in a change in solution pH. (b) Ionic liquid-gate FETs (α=159) were used to measure the change in solution pH as a function of the histone H1 concentration ([H1]), and thereby infer the activity of Cdk5 under steady-state conditions. (c) The FET measurements were in quantitative agreement with a complementary assay that used radioactively labeled γ-$^{32}$P-ATP as a reporter of Cdk5 activity. (d) Time-series measurements of enzyme catalyzed phosphorylation as a function of [H1] (9.1 μM, △; 12.7 μM, □; 18.2 μM, ○) are shown and allow the direct estimation of the reaction dynamics. The solid lines depict the reaction kinetics model that describes the time course of phosphorylation, while the dashed lines represent an estimate of the reaction velocity during the first 100 s after a change in the signal was detected.

We estimated the change in the solution pH for the FET measurements from $\Delta V_{t,BG}$ in Figure 4b (α=159). The expected change in the solution pH was estimated using the expression $\frac{d[H^+]}{d\,pH} =$ $-2.303 \frac{C_a K_a [H^+]}{(K_a+[H^+])^2}$, where $C_a$ is the buffer concentration, $K_a$ is the acid dissociation constant and $[H^+]$ is the proton concentration.[11,51] The change in pH was consistent with ≈3 phosphorylation sites on the substrate protein, assuming an electrolyte solution buffered with $C_a$=250 μM 3-(N-morpholino)propanesulfonic acid (MOPS).

**Enzyme Kinetics.** The FET-based measurements have a response time that allows the direct estimation of reaction kinetics and velocities as seen from Figure 4d and Figure S6 (see *Supplementary Information*). The concentration of the Cdk5/p25 complex in these measurements was 3.8 μM (5-fold lower than the quantity used in Figure 4b and c). A control sample without histone showed no change in the measured potential upon addition of ATP (*data not shown*). From Figure 4d, we observed that upon addition of ATP there was a decrease in $V_{BG}$ after ≈2 min. The polarity of $V_{BG}$ is consistent with the release of protons into solution, which results in an increase in $V_{LG}$. The initial reaction velocities were estimated from a linear regression of the first 100 s data after a change in $V_{BG}$ was detected and were found to increase monotonically with [H1]. Furthermore, the initial linear change in $V_{BG}$ is consistent with an enzyme limited reaction.[52] Finally, the time-course of each reaction in Figure 4d was fit with a first order rate law of the form, $V_{BG} = \beta(1 - e^{-k_1 t})$, where $\beta$ is a scaling constant and $k_1$ is a rate constant. The rate constant was consistent with previously reported values[52,53] and estimated to be $k_1$=0.18±0.02 per min, independent of the histone H1 concentration. FET-based measurements are in excellent quantitative agreement with existing techniques as seen from Figure 4, while enabling results in minutes.

**Conclusions.** We have demonstrated that dual-gate FETs with atomically thin channels and ionic liquid top-gates allow performance limited by the intrinsic quantum capacitance of the channel material, providing a robust and sensitive biosensing platform. The nearly 100-fold improvement in pH sensitivity above the Nernst value and order of magnitude improvement in the LOD over state-of-the-art silicon ISFETs at a bandwidth of 10 Hz allowed the measurement of small changes in solution pH during enzyme mediated phosphorylation under physiological conditions. The method eliminates the need for specialized labeling and hazardous material handling and can decrease the cost and complexity of assays appreciably. Another key advantage of the measurements is that they are time-resolved, enabling the estimation of both enzyme activity and kinetics in a single assay. Because signal transduction is performed remotely, the measurements are compatible with standard microtiter plates, and therefore amenable for use in high throughput screening to allow rapid evaluation of pharmaceutical candidates for neurological diseases.

The high sensitivity and SNR allowed measurements with 5-fold lower enzyme concentrations than physiological values and over an order of magnitude below previous demonstrations using FET. This in turn will enable new biomarker diagnostics from blood or cerebrospinal fluid to be developed for use at early stages of neurodegeneration where the changes to kinase activity are subtle and currently undetectable.[54,55] Importantly, these measurements can allow testing several years before any observable decline in cognitive function, allowing early interventions to be developed. The extended gate configuration of the sensors permit remote measurements, which allow them to be adapted for use in cell and tissue culture assays, where an increase in kinase activity has been correlated with pH imbalances.[56] Such measurements can allow assays that interrogate mechanistic behavior across multiple spatial scales – from test tube to cell culture experiments – to better quantify the origin of disease states.[57]

**Acknowledgements**
S.T.L. acknowledges support by the National Institute of Standards and Technology (NIST) grant 70NANB16H170. J.B.K. and N.B.G. acknowledge support by NIST grant 70NAHB15H023. A.C. acknowledges support by the NIST grant 70NANB17H259. Research performed in part at the NIST Center for Nanoscale Science and Technology nanofabrication facility.

## TOC Image

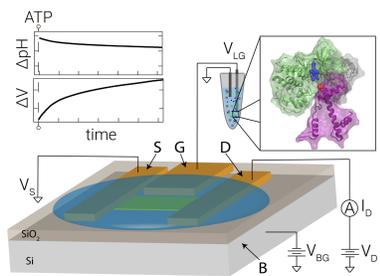